# Deep Learning and Image Super-Resolution-Guided Beam and Power Allocation for mmWave Networks

Yuwen Cao, Tomoaki Ohtsuki, *Senior Member, IEEE*, Setareh Maghsudi, and Tony Q. S. Quek, *Fellow, IEEE*

*Abstract*—In this paper, we develop a deep learning (DL)-guided hybrid beam and power allocation approach for multiuser millimeter-wave (mmWave) networks, which facilitates swift beamforming at the base station (BS). The following persisting challenges motivated our research: (i) User and vehicular mobility, as well as redundant beam-reselections in mmWave networks, degrade the efficiency; (ii) Due to the large beamforming dimension at the BS, the beamforming weights predicted by the cutting-edge DL-based methods often do not suit the channel distributions; (iii) Co-located user devices may cause a severe beam conflict, thus deteriorating system performance. To address the aforementioned challenges, we exploit the synergy of supervised learning and super-resolution technology to enable low-overhead beam- and power allocation. In the first step, we propose a method for beam-quality prediction. It is based on deep learning and explores the relationship between high- and low-resolution beam images (energy). Afterward, we develop a DL-based allocation approach, which enables high-accuracy beam and power allocation with only a portion of the available time-sequential low-resolution images. Theoretical and numerical results verify the effectiveness of our proposed framework.

## I. INTRODUCTION

Future wireless networks shall guarantee high data rates, energy- and spectral efficiency, low latency, and high reliability in numerous smart terminals for emerging applications. To achieve that goal, millimeter-wave (mmWave) communication with large-scale antenna arrays is a promising technology [1], [2]. That is because the mmWave spectrum is substantially wide so it enables multi Gigabit-per-second (Gbit/s) data rates [3], [4].

Despite several advantages, mmWave communication also suffers from some shortcomings. One of them is the severe path loss caused by the operation at high frequencies that allows a shorter wavelength than that of the conventional communication at lower frequency bands [5]. A method to address this challenge is performing beamforming and beam-steering with substantial antenna array gains [6]; However, fully digital beamforming architecture requires several energy-intensive radio-frequency (RF) chains, which dramatically increases the power consumption and hardware cost [5]. In contrast, analog beamforming is a viable option, as it requires only a single-RF chain, thus reducing the costs.

The work of Y. Cao was supported by Shanghai Sailing Program under Grant 23YF1400800. The work of S. Maghsudi was supported by Grant 16KISK035 from the German Federal Ministry of Education and Research (BMBF). The work of T. Q. S. Quek was supported by the National Research Foundation, Singapore and Infocomm Media Development Authority under its Future Communications Research & Development Programme.

By employing many antennas, one achieves large antenna array gains in mmWave communications; nevertheless, the excessive overhead of pilot training and rapid channel fluctuations make it challenging to estimate the channel accurately. References [7]–[10] develop solutions based on deep learning to tackle this challenge. In [7], *Rezaie et al.* propose a deep neural network (DNN)-based beam selection approach that exploits users' orientation and position. Similarly, the authors of work [8] use DNN architectures in multiple-input multiple-output (MIMO) systems to optimize the beamformer design with low-complexity benefits. In [9], the authors propose a DNN-based analog beam selection scheme via precisely estimating the received power using super-resolution. Reference [10] proposes offline learning-based beamforming approaches which enable fast adaption in a new wireless environment. There is a large body of research on deep learning-based beamforming prediction; Nonetheless, the predicted beamforming weights often do not suit the underlying channel distribution in practice, particularly with highly-mobile users. Besides, the existing DNN-based beam selection solutions are costly and invalid for multiuser mmWave networks. Hence, developing efficient beamforming methods that swiftly adapt to the evolving wireless environment is critical.

Against this background, we develop a method for downlink beam- and power allocation in a multiuser mmWave network. The objective is to facilitate swift beamforming at the base station (BS). We first formulate the downlink hybrid beam- and power allocation optimization problem. The problem is mix-integer programming and NP-hard. To tackle this challenge, we first design a high-resolution beam-quality prediction model. Besides, to enhance the prediction accuracy, we propose a *time-sequential low-resolution beam image dataset generation* framework. Afterward, we develop a deep learning-based beam- and power allocation approach that assigns beam and power to intended user pieces of equipment (UEs) with low overhead. The technical contributions of this paper are as follows:

- We develop a beam-image dataset generation and high-resolution beam-image prediction framework by exploiting deep learning and super-resolution networks (SRNs), to predict the missing or future beam domain qualities with low overhead.
- We propose a deep learning-based beam- and power allocation approach that completely avoids beam conflicts by picking only the top-$m$ preferred beams. We empirically identify the optimal value of $m$.
- The proposed allocation approach assigns the beams to

intended UEs and optimizes the downlink power allocation based on the predicted received data without requiring a knowledge of the underlying channel distribution.
- The proposed beamforming approach is not limited to the mmWave networks with mobile users, as it quickly maps the past observations into instantaneous beamforming vector with low training overhead.
- Simulation results show that our beam-quality prediction approach returns precise predictions with the least mean square error (MSE). Besides, compared with the existing allocation approaches, it enables near-optimal performance with low-overhead benefits.

## II. SYSTEM MODEL

We consider a mmWave communication system with a uniform planar array (UPA). The BS with an $M_{tx} = M_v \times M_h$ UPA transmits data signals to $K$ single-antenna UEs, where $M_v$ and $M_h$ respectively denote the number of antennas along the vertical- and horizontal axis.

### A. Channel Model

The BS serves the UEs with carrier frequency $f = 60$ GHz. For the channel, we adopt a geometry model of $L_p$ output paths for each channel realization. That includes the angle of arrival (AoA), angle of departure (AoD), path loss, path gain, etc. Thus the mmWave channel vector $\mathbf{h}_k \in \mathbb{C}^{M_{tx}}$ between the BS and the $k$th UE can be expressed as [1]

$$\mathbf{h}_k = \sqrt{\frac{M_{tx}}{L_p}} \sum_{l=1}^{L_p} \alpha_{k,l} \mathbf{a}(\varphi_{k,l}, \phi_{k,l}), \ \forall k \in \{1, \cdots, K\}, \quad (1)$$

in which $\mathbf{a}(\cdot)$ denotes the steering vector function. The steering vector of a UPA is given by $\mathbf{a}(\varphi_{k,l}, \phi_{k,l}) = \mathbf{a}^v(\phi_{k,l}) \otimes \mathbf{a}^h(\varphi_{k,l}, \phi_{k,l})$, where $\mathbf{a}^v(\phi_{k,l}) \in \mathbb{C}^{M_v}$ and $\mathbf{a}^h(\varphi_{k,l}, \phi_{k,l}) \in \mathbb{C}^{M_h}$ represent the steering vector whose elements lie in the vertical- and horizontal axis, respectively. $\otimes$ denotes the Kronecker product operation. In addition, $\varphi_{k,l}$ and $\phi_{k,l}$ are the AoD azimuth and elevation of the $l$th path of UE $k$, respectively. We consider a half-wavelength $\lambda/2$ spaced UPA at BS, so that [12]

$$\mathbf{a}^v(\phi_{k,l}) = \left[1, e^{-j\pi \cos(\phi_{k,l})}, \cdots, e^{-j\pi(M_v-1)\cos(\phi_{k,l})}\right]^T, \quad (2)$$

$$\mathbf{a}^h(\varphi_{k,l}, \phi_{k,l}) = \left[1, e^{-j\pi \sin(\phi_{k,l})\sin(\varphi_{k,l})}, \cdots, e^{-j\pi(M_h-1)\sin(\phi_{k,l})\sin(\varphi_{k,l})}\right]^T. \quad (3)$$

$\alpha_{k,l} = \frac{\sqrt{M_{tx}} \times \lambda \times g}{4\pi d_{k,l}^p} e^{\frac{-j2\pi d_{k,l}^p}{\lambda}}$ in (1) is the complex-valued gain of the $l$th path of UE $k$ [13], where $\lambda$ is the wavelength, $g$ is the reflection gain, and $d_{k,l}^p$ is the path distance. [2]

### B. Data Streaming

We define $\mathcal{W}$ as the standard discrete Fourier transform (DFT)-based codebook for the UPA-based transmitter. Within the pilot training phase, the UPA employs the DFT codebook-based analog beamforming architecture to beamform signals with a single-RF chain [7]. The analog beamforming weights associated with UE $k \in \{1, \cdots, K\}$ after applying to a UPA is given by $\mathbf{w}_{k,t} = \mathbf{w}_{k,t}^v \otimes \mathbf{w}_{k,t}^h$, where $t \in \{1, \cdots, \tau\}$ and $\tau$ is the number of time frames in the pilot training phase. $\mathbf{w}_{k,t}^v = [w_1^v, \cdots, w_{M_v}^v]^T$ and $\mathbf{w}_{k,t}^h = [w_1^h, \cdots, w_{M_h}^h]^T$ are the beamforming weights on antennas concerning vertical- and horizontal axis, respectively.

The BS first sweeps the beams by broadcasting them to the UEs in sequential time slots. After applying the analog beamformer $\mathbf{w}_{k,t} \in \mathcal{W}$, the UE $k$ observes the signal $\mathbf{r}_k = \sqrt{p_k}\mathbf{h}_k^H \mathbf{w}_{k,t} \mathbf{s} + \sum_{i \neq k} \sqrt{p_i}\mathbf{h}_k^H \mathbf{w}_{i,t} \mathbf{s} + \mathbf{n}_k$, where $p_k$, $\mathbf{s}$, and $\mathbf{n}_k$ respectively denote the transmit power budget for UE $k \in \{1, \cdots, K\}$, the known training signal with normalized power, and the zero-mean AWGN with variance $N_0$. Accordingly, the UE feeds back the channel measurements to BS through the mmWave control channel for subsequent beamforming weight prediction and channel estimation. For $\forall k \in \{1, \cdots, K\}$, the beamforming vector prior to time frame $t+1$ yields

$$\mathbf{w}_{k,t+1} = \mathcal{F}_t(\mathbf{w}_{k,1:t}, \mathbf{r}_{k,1:t}), \ \forall t \in \{1, \cdots, \tau\}, \quad (4)$$

where $\mathcal{F}_t(\cdot)$ is the beamforming strategy in time frame $t+1$ as a function of prior observations in terms of $\tau$ consecutive beamforming vectors and channel measurements.

Our proposed approach guarantees the following properties:
- In the spatial domain, only $m_{tx} = m_v \times m_h$ low-resolution beam images are essential inputs to the neural network that predicts the analog beamforming for each UE, i.e., $m_{tx} \ll M_{tx}$.[3] Thus the training overhead of our neural networks reduces with low-dimensional input. [4]
- In the time domain, only $s$ time-sequential low-resolution beam images are collected to feed neural networks with $s < t-1$, to further reduce the involved training overhead, thereby enabling lightweight neural networks. Details of the neural network architecture design appear in Sec. IV.

## III. PROBLEM FORMULATION

Our objective is to optimize the beam- and power allocation decisions at BS that achieves high sum-rate performance with low overhead. To this end, we define $\mathcal{R}(\{\mathbf{w}_{k,t}\}_{k=1}^K, \mathbf{p})$ as the sum-rate, i.e., $\mathcal{R}(\{\mathbf{w}_{k,t}\}_{k=1}^K, \mathbf{p}) = \sum_{k=1}^K \log_2(1 +$ $\gamma_k(\mathbf{w}_{k,t}, p_k))$, where $\mathbf{p} = [p_1, \cdots, p_K]^T$ is the downlink power allocation vector. $\gamma_k(\mathbf{w}_{k,t}, p_k)$ is the received signal-to-interference-plus-noise ratio (SINR) of UE $k$ associated with a downlink allocation policy $(\mathbf{w}_{k,t}, p_k)$ in time frame $t$, i.e.,

$$\gamma_k(\mathbf{w}_{k,t}, p_k) = \frac{p_k \|\mathbf{h}_k^H \mathcal{F}_{t-1}(\mathbf{w}_{k,1:t-1}, \mathbf{r}_{k,1:t-1})\|^2}{\sum_{i \neq k} p_i \|\mathbf{h}_k^H \mathcal{F}_{t-1}(\mathbf{w}_{i,1:t-1}, \mathbf{r}_{i,1:t-1})\|^2 + N_0}, \quad (5)$$

where $\|\cdot\|$ denotes the $l_2$-norm operation.

---

[1]We adopt a small-scale channel modeling for the pilot training phase, and ignore the *Doppler shifts* in this phase. This is because: i) our beamforming strategy depends on the observed information in terms of beamforming vectors and channel measurements, rather than the information of the channel distribution; ii) the Doppler shifts impose little influence on the beamforming strategy in our system modeling. In addition, we will study the deep learning-based beamforming with a fast-varying channel model (large Doppler shifts [11]) in our future works.

[2]In this paper, we generate a channel via a 3D ray-tracing simulator called as Wireless Insite [14]. The output of this simulator is a time-variant channel impulse response. For every coherence time, along with UE's movements, both the AoA and AoD of each link, the amount of mmWave links, and the mobile distance of UE vary, thus leading to the rapid changes of received signal strength (RSS). This is due to the high sensitivity of mmWave signal propagation to blockage [15].

[3]We reshape the low-resolution beam list of length $m_{tx}$ into a two-dimensional low-resolution beam array of size $m_v \times m_h$.

[4]Please notice that, although the proposed deep learning-based approach enables an overhead reduction, the pilot overhead increases proportionally with the number of pilot transmissions.

The binary matrix $\mathbf{U} \in \{0,1\}^{K \times N \times \tau}$ summarizes the beam assignment decisions: $u_{k,n,t} = 1$ means that beam $n$ shall be assigned to UE $k$ in time frame $t$ and zero otherwise. For all $n \in \{1, \cdots, M_{tx}\}$ and $t \in \{1, \cdots, \tau\}$, we have $\sum_{k=1}^K u_{k,n,t} \leq 1$. After applying this beam-assignment strategy at BS, we have $\tilde{\mathbf{w}}_i = \sum_{k=1}^K u_{k,n,t} \mathbf{w}_{k,t}$ associated with the possible combination $i = (k,n)$, $i \in \{1, \cdots, I\}$ and

$$I = \frac{(M_v \times M_h)!}{(M_v \times M_h - K)!}. \quad (6)$$

We formulate the joint beam-assignment and power allocation optimization problem (**P1**) as [5]

$$\underset{\mathbf{U}, \mathbf{p}, \mathcal{F}_t(\cdot)}{\text{maximize}} \quad \mathcal{R}\left(\{\tilde{\mathbf{w}}_i\}_{i=1}^I, \mathbf{p}\right) \quad (7a)$$

$$\text{s.t. } \mathbf{w}_{k,t+1} = \mathcal{F}_t(\mathbf{w}_{k,1:t}, \mathbf{r}_{k,1:t}), \forall k, t \in \{1, \cdots, \tau\}, \quad (7b)$$

$$\sum_{k=1}^K u_{k,n,t} \leq 1 \,\&\, u_{k,n,t} \in \{0,1\}, \,\forall n \in \{1, \cdots, M_{tx}\}, \quad (7c)$$

$$\sum_{k=1}^K p_k \leq P_{\max}. \quad (7d)$$

The problem **P1** is an NP-hard mix-integer programming problem, and finding its optimum $(\mathbf{U}, \mathbf{p})$ solution is involved due to the following reasons: (i) The solution space of the problem **P1** increases exponentially with the beamforming dimension $M_{tx}$ and the number of UEs $K$ so that the conventional solutions suffer from prohibitive computational complexity; (ii) Although the deep learning-based solution predicts the beamforming weights, in practice, such predicted beamforming weights do not suit the underlying channel distribution [10]; (iii) An accurate estimation of the best beam with the highest received power via DNN-based solutions [7], [9] is costly and invalid for multiuser mmWave communications. Therefore, it is imperative to develop an efficient beamforming approach $\mathcal{F}_t(\cdot)$ that can swiftly adapt to the evolving wireless environment. In addition, deep learning has achieved competitive performance in various tasks, especially in image processing [18], and object detection using images [19], [20]. This motivates us to propose a deep learning and image super-resolution-guided analog beamforming approach (Sec. V), based on the predicted data via the following high-resolution beam-quality image prediction framework (Sec. IV).

## IV. HIGH-RESOLUTION BEAM-QUALITY IMAGE PREDICTION

In the phase domain, the received power of uniformly placed wide/narrow beams can be viewed as a low/high-resolution beam-quality image. Thus we focus on increasing the *temporal and spatial resolution* of beam images by exploiting the supervised learning and SRNs. We use the time-sequential set of $s$ consecutive low-resolution beam-quality images with size $m_v \times m_h$ as input to predict the high-resolution beam-quality images of size $M_v \times M_h$. In other words, we cast the beam-quality image prediction problem as a *multi-output nonlinear regression* problem. To tackle such a challenge, we formulate the beamformer design given in (4) as a mapping from past observations, i.e.,

$$\mathbf{y}_{k,t+1} = \hat{\mathcal{F}}_t(\mathbf{r}_{k,t-s+1:t}), \,\forall s \in \{1, \cdots, t-1\}, \quad (8)$$

where $\hat{\mathcal{F}}_t(\cdot)$ is the beam-quality image (energy) $\mathbf{y}_{k,t+1}$ at time $t+1$ as a function of the past channel measurements.

*Low-resolution beam image dataset generation*: To reduce the training overhead and enable fast training per epoch of our beam-quality prediction module, we pick only a portion of $s$ beam images from experience to generate the time-sequential low-resolution data $\mathbf{X}$. Notably, $s$ is defined as another hyperparameter to optimize. To improve the prediction accuracy of our downlink beam-quality image prediction model, both the temporal and spatial correlations in the beam qualities are utilized with the 3D convolutional LSTM (3D Conv-LSTM) architecture.

*High-resolution beam quality image prediction*:
Essentially, predicting high-resolution beam image $\mathbf{y}$ is a procedure that explores the temporal and spatial correlations in the beam qualities using context information. The proposed high-resolution beam-quality image prediction module is shown in Fig. 1, which considers the 3D Conv-LSTM framework including the convolutional 2D (Conv2D) layer, the sub-pixel convolution 3D (SubPixel Conv3D) layer, the 3D ConvLSTM layer, the flatten (FL) layer, the fully connected (FC) dense layer, and the batch normalization (BN) layer.

We use the standard MSE as the loss function to measure the squared error loss of our neural networks. For a given batch size $Q$, the loss function is defined as

$$f(\mathbf{X}^{(i)}, \mathbf{y}^{(i)}; \theta) = \frac{1}{Q} \sum_{i=1}^Q \|\hat{\mathbf{y}}^{(i)}(\theta) - \mathbf{y}^{(i)}\|^2, \quad (9)$$

where $\theta$ is the network parameter. $\mathbf{y}^{(i)}$ and $\hat{\mathbf{y}}^{(i)}(\theta)$ are the high-resolution beam-quality image (ground truth image) and the predicted high-resolution beam-quality image, respectively.

## V. DOWNLINK BEAM AND POWER ALLOCATION POLICY

### A. Optimal Allocation Policy

Note that the beam allocation in principle is the codeword allocation. The optimal beam allocation strategy is to calculate the sum rate in (7) for finding out $K$ different codewords exhaustively from the DFT-based codebook $\mathcal{W}$. As such, the beam associated with the largest sum rate will be selected and assigned to the intended UE. To be concrete, the proposed algorithm is constituted by the following three major stages.

- **Beam-allocation update**: The total possible permutation of picking $K$ different beams from DFT-based codebook $\mathcal{W}$ is $I = \frac{M_{tx}!}{(M_{tx}-K)!}$. For the $i$th iteration, $i \in \{1, \cdots, I\}$, we pick one possible combination $\mathcal{W}_i = \{\tilde{\mathbf{w}}_{i_1}, \cdots, \tilde{\mathbf{w}}_{i_K}\}$ as one candidate beam set for $K$ UEs.

---

[5]Problem (**P1**) describes joint beam assignment and power allocation optimization in the mmWave downlink communication system. Besides, the general downlink beamforming optimization problems defined in [16] and [17] are polynomial-time reducible to optimization problem (**P1**). Therefore, problem (**P1**) is NP-hard.

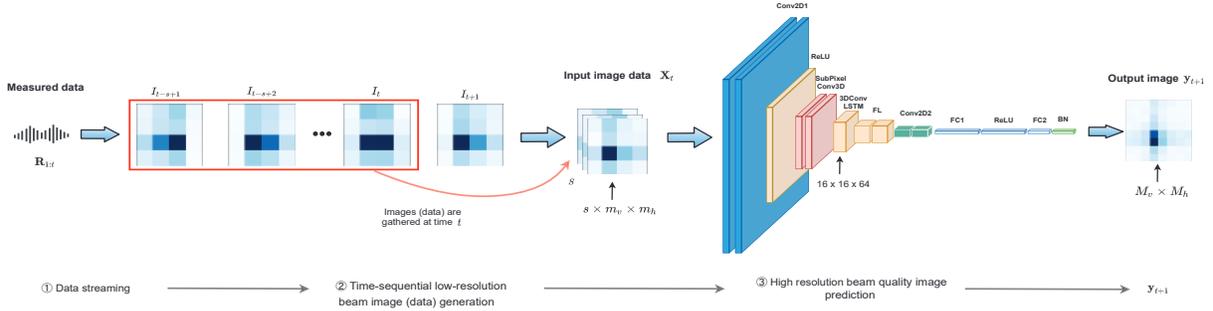

Fig. 1. An overview of the proposed high-resolution beam-quality image prediction framework.

- **Power-allocation update**: For the $i$th iteration, we update the downlink power allocation vector $\mathbf{p}_i$ by addressing the constrained sum-rate maximization problem of

$$\underset{\mathbf{p}_i}{\text{maximize}} \ \mathcal{R}(\tilde{\mathbf{w}}_{i_1}, \cdots, \tilde{\mathbf{w}}_{i_K}, \mathbf{p}_i) \quad \text{s.t.} \sum_{k=1}^{K} p_{i,k} \leq P_{\max}. \quad (10)$$

We next introduce the following Lagrangian function

$$\mathcal{L}(\mathbf{p}_i, \mu) = \mathcal{R}(\tilde{\mathbf{w}}_{i_1}, \cdots, \tilde{\mathbf{w}}_{i_K}, \mathbf{p}_i) - \mu\left(\sum_{k=1}^{K} p_{i,k} - P_{\max}\right), \quad (11)$$

in which $\mu > 0$ indicates a Lagrangian multiplier [21]. Since the channel information is assumed to be unknown in prior, estimating $\|\mathbf{h}_k^H \tilde{\mathbf{w}}_{i_k}\|^2$ seems to be rather difficult. Instead, we approximate $\|\mathbf{h}_k^H \tilde{\mathbf{w}}_{i_k}\|^2$ by applying the SRNs and least-square (LS) estimator [22], based on the predicted received data and known training signal. To be concrete, we first predict the received power in real-part $\hat{\mathbf{R}}_k^R \in \mathbb{R}^{M_v \times M_h}$ and the received power in imaginary-part $\hat{\mathbf{R}}_k^I \in \mathbb{R}^{M_v \times M_h}$ by exploiting SRNs, thus yielding $\hat{\mathbf{R}}_k = \sqrt{\hat{\mathbf{R}}_k^R} + j\sqrt{\hat{\mathbf{R}}_k^I}$. Let $\hat{\mathbf{h}}_k$ be the estimated channel between BS and UE $k$. Then we have $\text{vec}(\hat{\mathbf{R}}_k) = \sqrt{p_k}\hat{\mathbf{h}}_k^H \tilde{\mathbf{w}}_{i_k} \mathbf{s} + \mathbf{n}_k$, where $p_k$ denotes the transmit power preseted in beam-sweeping stage and $p_k > 0$. Afterwards, based on this predicted $\hat{\mathbf{R}}_k \in \mathbb{C}^{M_v \times M_h}$ and by applying the LS estimator [22], we have

$$\hat{\mathbf{h}}_k^H \tilde{\mathbf{w}}_{i_k} = \frac{\mathbf{s}^\dagger \times \text{vec}(\hat{\mathbf{R}}_k)}{\sqrt{p_k}}, \quad (12)$$

where $(\cdot)^\dagger$ represents the pseudo-inversion of a vector. In such a case, we address the power allocation optimization problem in (10) by means of the Karush-Kuhn Tucker (KKT) conditions [21]. From $\frac{\partial \mathcal{L}(\mathbf{p}_i, \mu)}{\partial p_{i,k}} = 0$, we have

$$p_{i,k}^* = \max\left(\frac{1}{\mu^*} - \frac{\sum_{j \neq k} p_{i,j}\|\hat{\mathbf{h}}_k^H \tilde{\mathbf{w}}_{i_j}\|^2 + N_0}{\|\hat{\mathbf{h}}_k^H \tilde{\mathbf{w}}_{i_k}\|^2}, 0\right). \quad (13)$$

In addition, $\mu^*$ can be obtained by exploiting the bi-section search method under the other KKT condition $\sum_{k=1}^{K} p_{i,k} \leq P_{\max}$. In this way, the updated power allocation vector $\mathbf{p}_i = [p_{i,1}^*, \cdots, p_{i,K}^*]^T$ can be attained.

- **Joint-allocation policy determination**: Based on the above two stages, we can obtain the combination of $(\tilde{\mathbf{w}}_{i_1}, \cdots, \tilde{\mathbf{w}}_{i_K}, \mathbf{p}_i)$ for each iteration of our algorithm.

Accordingly, we can achieve the optimal beam and power allocation policy by solving

$$(\mathbf{U}^*, \mathbf{p}^*) = \arg\max_{i \in \{1, \cdots, I\}} \mathcal{R}(\tilde{\mathbf{w}}_{i_1}, \cdots, \tilde{\mathbf{w}}_{i_K}, \mathbf{p}_i). \quad (14)$$

Notably, the above approach provides a guarantee of the following property.

**Proposition V.1.** The power allocation policy stated above can be interpreted as a procedure that each time frame maps time-sequential low-resolution beam images (i.e., energy) $\mathbf{R}_{t-s+1:t}$ into a downlink power allocation vector $\mathcal{G}_t(\mathbf{R}_{t-s+1:t}) = \{p_1^*, \cdots, p_K^*\}$, $s \in \{1, \cdots, t-1\}$, by exploiting the LS and SRNs. Besides, since $\mathcal{G}_t(\mathbf{R}_{t-s+1:t})$ shall subject to the power budget constraint in (10), we, therefore, deduce that the power allocation vector set

$$\mathbf{P} = \{\mathcal{G}_t(\mathbf{R}_{t-s+1:t}) : \mathbf{R} \in \mathbb{R}^{m_v \times m_h \times 2}\}, \quad (15)$$

in principle is a convex set.[6]

*Proof*: See Appendix A in [23]. ∎

### B. Deep Learning-Based Allocation Policy

To further reduce the overhead incurred for allocating the best beam and power in the aforementioned solution, we develop a sub-optimal allocation approach by using deep learning. To be concrete, we first invoke the beam-quality prediction approach to predict the high-resolution image and thus yielding a set of beam-quality lists. Such a beam-quality list classifies the $M_{tx}$ beams in descending order from the strongest received power to the lowest received power. In this way, we assign the strongest beam for each UE as the preferred beam. However, for multiuser mmWave systems, some UEs' locations may be co-located together, thus rendering that some of them may share the same beam, thus causing severe beam conflicts. As such, we find out those UEs that have the unique strongest beams based on their beam-quality lists. For ease of simplicity, we assume that there have $\gamma$ UEs whose strongest beams are different, meaning that $K - \gamma$ UEs share the same one or multiple strongest beams.[7] Subsequently, we solve the beam allocation problem against those $K - \gamma$ UEs by picking only top-$m$ ranked beams from their beam-quality lists.

---
[6]The superscript 2 hereby refers to two types of low-resolution images (energy).

[7]Such as, a portion of the $K - \gamma$ UEs as a subset share the same beam $l_s$, while the rest of the $K - \gamma$ UEs as another subset share the same beam $l_t$ with $l_s, l_t \in \{1, \cdots, M_{tx}\}$ and $l_s \neq l_t$.

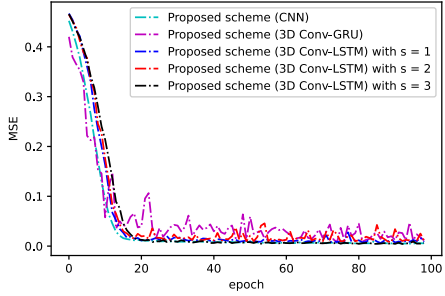

Fig. 2. MSE performance of the proposed scheme employing the 3D Conv-LSTM architecture under various hyperparameter $s = \{1, 2, 3\}$ compared to the state-of-the-art neural network architectures.

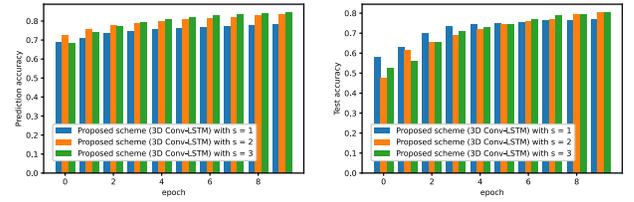

(a) Prediction accuracy    (b) Test accuracy

Fig. 3. Accuracy of our proposed scheme employing the 3D Conv-LSTM architecture under various hyperparameter $s = \{1, 2, 3\}$.

**Proposition V.2.** It is noteworthy that, the corresponding probability of selecting $K$ different codewords from top-$m$ ranked beams without beam conflicts becomes

$$\mathbb{P}(m, \gamma) := \frac{m!}{m^{K-\gamma}(m-K+\gamma)!}, \quad K - \gamma \leq m \ll M_{tx}. \quad (16)$$

*Proof*: Please see Appendix B in [23]. ∎

Based on the aforementioned analyses, to allocate $K - \gamma$ distinct codewords from top-$m$ beams without beam conflicts, we, therefore, develop the following joint beam-assignment and power allocation optimization problem (**P2**):

$$\underset{\mathbf{U}, \mathbf{p}}{\text{maximize}} \quad \mathcal{R}\big(\{\tilde{\mathbf{w}}_{i'}\}_{i'=1}^{I'}, \mathbf{p}\big) \quad (17a)$$

$$\text{s.t.} \sum_{k'=1}^{K-\gamma} u_{k',n',t} \leq 1 \ \& \ u_{k',n',t} \in \{0, 1\}, \ \forall t, n' \in \{1, \cdots, m\}, \quad (17b)$$

$$\sum_{k=1}^{K} p_k \leq P_{\max}, \quad (17c)$$

where $I' = \frac{m!}{(m-K+\gamma)!}$ refers to the total possible permutation of picking $K - \gamma$ distinct beams from top-$m$ ranked beams.

The proposed approach selects the beam combination among $\frac{m!}{(m-K+\gamma)!}$ candidates. In addition, although re-evaluating the neural network model and re-predicting the high-resolution beam qualities of a set of $K$ UEs adds some complexity, such complexity is much lower than that of the candidate beam combination selection. Hence, the computational complexity of our approach yields $\mathcal{O}\left(\frac{m!}{(m-K+\gamma)!}\right)$.

TABLE I. THE SIMULATION PARAMETER SETTING.

| Parameter description | Value |
|---|---|
| Carrier frequency $f$ | 60 GHz |
| Effective Bandwidth $B$ | 100 MHz |
| No. of antennas at BS $M_v \times M_h$ | $8 \times 8$ |
| No. of high (low) resolution beams | $8 \times 8$ ($4 \times 4$) |
| Training interval | 0.1 s |
| Height of BS | 10 m |
| No. of paths $L_p$ | 25 |
| Ray spacing (degrees) | $5 \times 10^{-4}$m |
| Total downlink power $P_{\max}$ | [-10 dBm, 12 dBm] |
| Signal-to-interference power ratio | 10 dB |
| Reflection gain $g$ | -6 dB |
| Noise figure | 9.5 dB |

## VI. PERFORMANCE EVALUATIONS

### A. Simulation Setup

We evaluate the performance of our beam-quality prediction module under the 3D Conv-LSTM, the convolutional neural network (CNN), and the 3D convolutional gated recurrent unit (3D Conv-GRU) architectures.[8] Besides, in our simulation the total params#,[9] the trainable params#, and the non-trainable params# are respectively to be <351,416, 351,160, 256>; the initial learning rate is $1 \times 10^{-3}$; the batch size $Q$ is 1024; and the neural network optimizer is the *Adadelta* algorithm. We adopt the Keras with the TensorFlow backend for implementing the aforementioned neural networks. Our simulations are based on the Wireless Insite [14]. **TABLE I** lists the mainly used simulation parameters which are set up by following the 3GPP NR framework [24]. In our experiment, the number of each UE's movements and the number of experiments are 29 and 2020, respectively. It is noteworthy that: i) the existing beam selection solutions [25], [26] are costly and invalid for the mmWave networks with mobile users; ii) the DNN-based beamforming prediction methods [7], [9] perform poorly in multiuser mmWave networks, particularly with high-mobility users. Nevertheless, in this paper, the proposed beamforming approach is not limited to the mmWave networks with mobile users, as it swiftly maps the past observations into beamforming vector with low training overhead. Besides, we will study the impact of mobility characteristics on the sum-rate performance in multiuser mmWave networks in our future works.

### B. Simulation Results

**Fig. 2** shows the MSE of the high-resolution beam quality $\mathbf{y}^{(i)}$ and the predicted $\hat{\mathbf{y}}^{(i)}(\theta)$ using our prediction approach. **Fig. 2** reveals that our approach employing the CNN can converge faster than the proposed approach using other deep learning algorithms. The reason is that CNN directly predicts the beam quality based on the simultaneously received power from UEs, thus facilitating faster training per epoch.

**Fig. 3(a)** and **Fig. 3(b)** illustrate the prediction accuracy and test accuracy of our beam-quality prediction module, respec-tively. As can be seen from **Fig. 3(a)** and **Fig. 3(b)**, our prediction approach employing the 3D Conv-LSTM architecture can provide high prediction accuracy and test accuracy. This is because our beam-quality prediction module incorporating the SRNs can precisely predict the beam qualities from prior experience via extracting the temporal changes.

---

[8]We adopt CNN to predict the high-resolution beam quality $\hat{\mathbf{y}}^{(i)}(\theta)$ at the current coherence time, based on the power received simultaneously from UEs. Besides, both 3D Conv-LSTM and 3D Conv-GRU are also adopted as the training structure for 3D images to capture spatiotemporal correlations based on the previous beam-quality measurements.

[9]By params#, we mean the number of parameters.

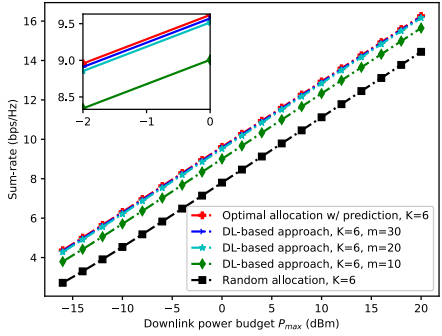

Fig. 4. Sum-rate performance of the proposed deep learning-based approach compared to the state-of-the-art with predicted channel information.

**Fig. 4** depicts the sum-rate performance of different beam and power allocation approaches with the *predicted* channel information. It is easy to see from Fig. 4 that our deep learning-based approach can provide almost the same sum-rate performance as that of the optimal allocation approach. Besides, Fig. 4 reveals that increasing the value of $m$ from 10 to 30 leads to sum-rate performance improvement. This is because increasing the value of $m$ can (i) reduce the beam confliction probability as stated in **TABLE I**, and (ii) enlarge the size of the testing set, thus making our neural network model better. The choice of $m$ is critical for the deep learning-based solution, which in fact belongs to an accuracy and complexity trade-off.

## VII. Conclusion

This paper developed a novel deep learning-guided downlink beam and power allocation approach for multiuser mmWave networks for facilitating swift beamforming at BS. Specifically, we first proposed a lightweight high-resolution beam-quality image prediction approach aiming at predicting beamforming weights from prior observations. Subsequently, we developed a deep learning-based allocation approach that can assign the desired beam and power to UEs without beam conflicts. Simulation results show that our approach enables sub-optimal performance with a low-overhead benefit.